\documentclass[journal]{defs/IEEEtran}
\IEEEoverridecommandlockouts
\usepackage{float}

\usepackage{cite}
\usepackage{amsmath,amssymb,amsfonts}
\usepackage{mathtools}
\usepackage{amsthm}
\usepackage{algorithm}
\usepackage{algpseudocode} 
\usepackage{etoolbox}
\usepackage{graphicx,color}
\usepackage{textcomp}
\usepackage[dvipsnames,rgb]{xcolor}
\usepackage{makecell}
\usepackage{float}
\usepackage{tikz}
\usetikzlibrary{arrows.meta, positioning, shapes.geometric,decorations.pathreplacing}
\usetikzlibrary{patterns}
\usepackage{siunitx}
\usepackage{tabularx}
\usepackage{colortbl}
\usepackage{multirow}
\usepackage{array}
\usepackage{acronym}

\usepackage{subcaption}
\usepackage{booktabs}
\usepackage{hyperref} 
\usepackage[capitalize,noabbrev]{cleveref}
\usepackage[textsize=tiny]{todonotes}
\usepackage{latexsym}
\usepackage{gensymb}
\usepackage{enumitem}
\usepackage{esvect}
\usepackage{comment}
\usepackage[normalem]{ulem}
\usepackage{soul}

\input{defs/macros}
\theoremstyle{plain}

\theoremstyle{definition}

\theoremstyle{remark}

\definecolor{nyuviolet}{RGB}{87, 6, 140}

\def\BibTeX{{\rm B\kern-.05em{\sc i\kern-.025em b}\kern-.08em
    T\kern-.1667em\lower.7ex\hbox{E}\kern-.125emX}}

\newacro{ACDD}{Alamouti with cyclic delay diversity}
\newacro{URLLC}{ultra-reliable low-latency communications}
\newacro{3GPP}{third generation partnership project}
\newacro{PHY}{physical layer}
\newacro{MIMO}{multiple-input multiple-output}
\newacro{MU-MIMO}{multi-user multiple-input multiple-output}
\newacro{SIMO}{single-input multiple-output}
\newacro{MISO}{multiple-input single-output}
\newacro{SISO}{single-input single-output}
\newacro{MRC}{maximum-ratio combining}
\newacro{SNR}{signal-to-noise ratio}
\newacro{CP}{cyclic prefix}
\newacro{CDD}{cyclic delay diversity}
\newacro{FSC}{frequency-selective channel}
\newacro{STC}{space-time coding}
\newacro{FFT}{fast Fourier transform}
\newacro{LMMSE}{linear minimum mean-squared error}
\newacro{CFAC}{cross frequency AoA consistency}
\newacro{FER}{frame error rate}
\newacro{OFDM}{orthogonal frequency division multiplexing}
\newacro{OCDM}{orthogonal chirp division multiplexing}
\newacro{RMS}{root mean square}
\newacro{DS}{delay spread}
\newacro{FSC}{frequency-selective channel}
\newacro{CSI}{channel state information}
\newacro{LMMSE-PIC}{linear minimum mean squared error with parallel interference cancellation}
\newacro{PFE}{perfect-feedback equalizer}
\newacro{FD}{full-duplex}
\newacro{PDP}{power delay profile}
\newacro{PDF}{probability density function}
\newacro{DFT}{discrete Fourier transform}
\newacro{SDFT}{sparse DFT}
\newacro{ICI}{inter-carrier interference}
\newacro{OTFS}{orthogonal time frequency space}
\newacro{AWGN}{additive white Gaussian noise}
\newacro{SWH}{sparse Walsh-Hadamard}
\newacro{LLR}{log-likelihood ratio}
\newacro{PMF}{probability mass function}
\newacro{CRC}{cyclic redundancy check}
\newacro{PAM}{pulse amplitude modulation}
\newacro{QAM}{quadrature amplitude modulation}
\newacro{FWHT}{fast Walsh-Hadamard transform}
\newacro{MAP}{maximum a-posteriori}
\newacro{SC}{specular component}
\newacro{CFO}{carrier frequency offset}
\newacro{ISI}{inter-symbol interference}
\newacro{ZP}{zero-padding}
\newacro{EVD}{eigenvalue decomposition}
\newacro{BCJR}{Bahl, Cocke, Jelinek, and Raviv}
\newacro{WHT}{Walsh-Hadamard transform}
\newacro{APP}{a-posteriori probability}
\newacro{SILE-EPIC}{self-iterated linear equalizer with expectation propagation}
\newacro{EP}{expectation propagation}
\newacro{i.i.d.}{independent and identically distributed}
\newacro{CWCU}{component wise conditionally unbiased}
\newacro{MSE}{mean squared error}
\newacro{EXIT}{extrinsic information transfer}
\newacro{MI}{mutual information}
\newacro{PAPR}{peak-to-average power ratio}
\newacro{DFT-s}{discrete Fourier transform-spread}
\newacro{AMP}{approximate message passing}
\newacro{GAMP}{generalized \ac{AMP}}
\newacro{VAMP}{vector \ac{AMP}}
\newacro{RSC}{recursive systematic convolutional}
\newacro{QPSK}{quadrature phase-shift keying}
\newacro{CFAR}{constant false alarm rate}
\newacro{PD}{probability of detection}
\newacro{PFA}{probability of false alarm}
\newacro{RV}{random variable}
\newacro{CDF}{cumulative distribution function}
\newacro{HD-ZP}{half-duplex ZP}
\newacro{FD-CP}{full-duplex ZP}
\newacro{DFRC}{dual-function radar communication}
\newacro{SINR}{signal-to-interference noise ratio}
\newacro{ISAC}{integrated sensing and communication}
\newacro{SI}{self-interference}
\newacro{RSI}{residual self-interference}
\newacro{ADC}{analog-to-digital converter}
\newacro{DAC}{digital-to-analog converter}
\newacro{ED}{energy-detection}
\newacro{IDFT}{inverse discrete Fourier Transform}
\newacro{SFFT}{symplectic finite Fourier transform }
\newacro{CRB}{Cram{\'{e}}r-Rao bound}
\newacro{ZC}{Zadoff-Chu}
\newacro{RMSE}{root mean square error}
\newacro{MMSE}{minimum mean-square error}
\newacro{UW}{unique word}
\newacro{GFDM}{generalized frequency division multiplexing}
\newacro{RRC}{root-raised cosine}
\newacro{UB}{upper bound}
\newacro{CEF}{channel estimation field}
\newacro{TRX}{transceiver}
\newacro{IF}{intermediate frequency}
\newacro{RF}{radio frequency}
\newacro{FPGA}{field programmable gate arrays}
\newacro{SDR}{software-defined radio}
\newacro{UWB}{ultra wideband}
\newacro{FR3}{frequency range 3}
\newacro{PCB}{printed circuit board}
\newacro{SMA}{SubMiniature version A}
\newacro{MUSIC}{multiple signal classification}
\newacro{CIR}{channel impulse response}
\newacro{FR}{Frequency Range}
\newacro{mmWave}{millimeter wave}
\newacro{LoS}{line-of-sight}
\newacro{AoD}{angle-of-departure}
\newacro{ESNR}{estimation SNR}
\newacro{AoA}{angle-of-arrival}
\newacro{SDNR}{signal-to-DMC-noise ratio}
\newacro{ULA}{uniform linear array}
\newacro{DMC}{dense multipath component}
\newacro{ML}{maximum-likelihood}
\newacro{IFFT}{inverse fast Fourier transform}
\newacro{LM}{Levenberg-Marquardt}
\newacro{ACF}{autocorrelation function}
\newacro{UWB}{ultra-wideband}
\newacro{SLAM}{simultaneous localization and mapping}
\newacro{STO}{sampling time offset}
\newacro{GLRT}{generalized likelihood ratio test}
\newacro{FD-ED}{frequency-domain energy detectors}
\newacro{IOU}{intersection over union}
\newacro{FLOP}{floating point operation}
\newacro{FLOPS}{floating point operations per second}
\newacro{LTBF}{long-term beamforming}
\newacro{UE}{user equipment}
\newacro{SRS}{sounding reference signal}
\newacro{BW}{bandwidth}
\newacro{RB}{resource block}
\newacro{RE}{resource element}
\newacro{BS}{base station}
\newacro{CG}{conjugate-gradient}
\newacro{SRAM}{static random access memory}
\newacro{DSP}{digital signal processor}
\newacro{VLSI}{very large-scale integration}
\newacro{NR}{New Radio}
\newacro{ASIC}{application-specific integrated circuit}
\newacro{gNB}{next-generation NodeB}
\newacro{SCS}{subcarrier spacing}
\newacro{WMMSE}{weighted minimum mean square error}
\newacro{WSR}{weighted sum-rate}
\newacro{ZF}{zero-forcing}
\newacro{UMB}{upper mid-band}
\newacro{DM-RS}{Demodulation Reference Signal}

\begin{document}

\title{Interference Suppression for \\ Massive MU-MIMO Long-Term Beamforming \\ with Matrix Inversion Approximation
}

\author{
Amirreza Kiani$^{\diamond}$,
Ali Rasteh$^*$,
Marco Mezzavilla$^{\diamond}$,
and Sundeep Rangan$^*$ \\
    $^\diamond$Dipartimento di Elettronica, Informazione e Bioingegneria (DEIB), Politecnico di Milano, Milan, Italy\\
    
	$^*$NYU WIRELESS, NYU Tandon School of Engineering, New York, USA \\
	Email: amirreza.kiani@polimi.it, ar7655@nyu.edu, marco.mezzavilla@polimi.it, srangan@nyu.edu 

}

\maketitle

\thispagestyle{empty}
\pagestyle{empty}

\acresetall
\begin{abstract}
Long-term beamforming (LTBF) is a widely-used scalable alternative to instantaneous multi-user MIMO processing that leverages slowly varying spatial channel statistics.  VLSI implementations require matrix inversion that become computationally challenging for massive MIMO systems with large number of antennas.   In this work, we show that dominant interferers significantly degrade the numerical conditioning of the LTBF covariance matrix, leading to severe performance loss in finite-precision implementations of polynomial and conjugate gradient (CG) based inversion methods. To address this issue, we propose a subspace nulling approach that operates solely on long-term channel statistics and acts as an implicit preconditioning step for LTBF. By projecting the received signal onto the orthogonal complement of the dominant interference subspace, the proposed method reduces the eigenvalue spread of the covariance matrix and improves numerical stability. Through ray-tracing simulations in a realistic 5G scenario, we demonstrate that the proposed method substantially reduces the number of CG iterations required to achieve near-optimal performance across floating-point and fixed-point implementations while preserving the low-overhead nature of LTBF.
\end{abstract}

 \section{Introduction}
Massive \ac{MIMO} has been a key enabler of capacity gains in 5G systems \cite{larsson2014massive,jin2023massive}. There is now growing interest in further scaling the number of antenna elements \cite{nokia2025massiveMIMO}. Beyond improving spectral efficiency, high-dimensional arrays offer significant gains in interference suppression \cite{jia2025joint} and enable operation over wider bandwidths \cite{akrout2023bandwidth}. However, scaling massive \ac{MIMO} to very large dimensions introduces substantial implementation challenges \cite{dai2021scalable}. \Ac{LTBF} \cite{lozano2007long} has emerged as a promising approach to address these issues by leveraging slowly varying channel statistics. In particular, \ac{LTBF} reduces channel estimation overhead and computational complexity, facilitating scalable and hardware-efficient implementations.


In a cellular MU-MIMO uplink, one computational bottleneck in \ac{LTBF} is the
need to compute an $N_{rx} \times N_{rx}$ matrix inverse
where $N_{rx}$ is the number of base station antennas
\cite{rasteh2025scalable}.
For large numbers of antennas, this computation must be
done in hardware that necessitates approximations.  Reduced precision methods as well as iterative algorithms such as conjugate gradient (CG)
with limited number of iterations are commonly used in such MIMO settings \cite{yin2015vlsi,dai2021scalable}.

A critical challenge in such approximate methods is 
the sensitivity to strong interferers.  Approximate methods tend to have low dynamic range that can degrade when one source arrives at high power. Standard cellular systems use uplink power control to limit the dynamic range.  However, in newer deployments with spectrum sharing and malicious jamming can lead to the presence of strong interference.  In these case, the use of low-precision arithmetic makes matrix inversion more challenging  \cite{xue2016fast}.

The main contributions of this paper are summarized as follows:
\begin{itemize}

\item \textbf{Finite-precision sensitivity of LTBF:}
We show that strong interference significantly degrades the accuracy and convergence of hardware-friendly matrix inversion methods, such as polynomial approximation and conjugate gradient, with the effect being particularly pronounced under low-precision implementations.

\item \textbf{Interference nulling as implicit preconditioning:}
We propose a subspace projection method based solely on long-term channel statistics to suppress dominant interference prior to \ac{LTBF} processing. We demonstrate that this operation improves the numerical conditioning of the covariance matrix, effectively acting as an implicit preconditioner for iterative inversion methods.

\item \textbf{Robust low-precision implementation:}
Through simulations, we show that the proposed approach substantially reduces the number of \ac{CG} iterations required to achieve near-optimal performance in FP32 and fixed-point implementations, thereby mitigating numerical instability induced by strong interference.


\end{itemize}

 \section{System Model}
\label{sec:sys-model}

We consider a multi-user uplink scenario in which the received signal is corrupted by a dominant interferer whose spatial signatures vary slowly over time and occupy a low-dimensional subspace. The received signal at subcarrier $n$ and \ac{OFDM} symbol $k$ is given by
\begin{equation}\label{eq:system}
\bs{y}[n,k] = \sum_{i} \bs{H}_i[n,k] \bs{x}_i[n,k] + \sum_{j \neq i} \bs{v}_j[n,k] + \bs{w}[n,k],
\end{equation}
where $N_{\mathrm{rx}}$ is the number of receiver antennas, $\bs{y}[n,k] \in \mathbb{C}^{N_{\mathrm{rx}}}$ is the received signal vector, $\bs{H}_i[n,k] \in \mathbb{C}^{N_{\mathrm{rx}} \times N_s}$ denotes the channel matrix associated with \ac{UE} $i$, $\bs{x}_i[n,k] \in \mathbb{C}^{N_s}$ is the transmitted symbol vector, and $\bs{w}[n,k]$ is the additive noise. The channel matrix $\bs{H}_i[n,k]$ is assumed to include any transmit-side precoding applied at \ac{UE} $i$. 

Let $\mc{E}_x$ denote the transmit energy per \ac{UE} per symbol, and assume
\begin{equation} \label{eq:xvar}
    \mathbb{E}\!\left[ \bs{x}_i[n,k]\bs{x}_i^H[n,k]\right] = 
    \frac{\mc{E}_x}{N_s} \bs{I}.
\end{equation}
In this formulation, $\bs{H}_i[n,k]\bs{x}_i[n,k]$ represents the desired signal of user $i$, while $\bs{v}_j[n,k] = \bs{H}_j[n,k]\bs{x}_j[n,k]$ denotes the contribution of the interferer $j$.

We assume that the interference lies in a low-rank subspace of the receive signal space and can therefore be suppressed by projecting onto the orthogonal complement of $\mathcal{R}(\bs{H}_v)$. Two practical approaches can be used to estimate this interference subspace:

\begin{itemize}
    \item \textbf{Non-coherent estimation:} When the interfering signals are unknown, the receiver forms a spatial covariance estimate
    \begin{equation}\label{eq:noncoh}
        \bs{Q}_v := \mathbb{E} \left[ \bs{v}[n,k] \bs{v}^H[n,k]\right],
    \end{equation}
    where the expectation is taken over an interval in which large-scale channel parameters remain approximately constant.

    \item \textbf{Coherent estimation:} When the interfering signals are known, matched filtering can be used to obtain channel estimates $\wh{\bs{h}}_j[n,k]$. Due to phase variations across time and frequency, non-coherent averaging is still required, leading to
    \begin{equation}\label{eq:Qv}
        \bs{Q}_v := \sum_j \mathbb{E} \left[ \wh{\bs{h}}_j[n,k] \wh{\bs{h}}_j^H[n,k]\right].
    \end{equation}
\end{itemize}

Both approaches rely on long-term averaging. In this work, we focus on the non-coherent estimation case while leaving the coherent approach for future investigation.



\section{Review of Multi-user Long-term Beamforming}
The key idea in multi-user long-term beamforming is to project the received signal $\bs{y}[n,k]$ onto a low-dimensional subspace that approximately suppresses the interference components $\bs{v}_j[n,k]$ arising from other users. Specifically, for each user $i$, the received signal is projected as
\begin{equation} \label{eq:zproj}
    \bs{z}_i[n,k] = \bs{G}_i \bs{y}[n,k],
\end{equation}
where $\bs{G}_i$ is an $r \times N_{\subsf rx}$ that maps
 the RX signal to some $r$-dimensional space for some $r < N_{\subsf rx}$. 
The projection matrix
is held constant over a long-period and is independent of the small-scale fading.  

To construct the projection matrix $\bs{G}_i$, we define the spatial covariance matrix of the channel corresponding to user $i$ as
\begin{equation}   \label{eq:Qjdef}
    \bs{Q}_i := \mathbb{E}\left[ 
        \bs{H}_i[n,k]\bs{H}_i[n,k]\herm \right],
\end{equation}
where the expectation is taken over the small-scale fading while assuming that the large-scale propagation parameters remain constant over the long-term estimation interval. In alignment with 5G \ac{NR} standards, long-term channel estimation can be performed using $N_{\subsf SRS}$ reference signals per \ac{UE}, distributed across the assigned resource blocks as:
 \begin{equation} \label{eq:srs}
     {\bs{Q}}_j = \frac{1}{N_{\subsf SRS}} {\bs{H}}_j{\bs{H}}_j\herm.
\end{equation}
The aggregate interference-plus-noise covariance matrix is then given by
\begin{equation} \label{eq:Qdef}
    \bs{Q} := \bs{I} + \sum_{i=1}^{N_{\rm UE}} \alpha_i \bs{Q}_i,
\end{equation}
where $\alpha_j$ denotes the effective transmit \ac{SNR} of user $j$. 
Under this formulation, the projection matrix that maximizes a capacity upper bound can be constructed as
\begin{equation} \label{eq:Giopt}
    \bs{G}_i = \left[\bs{Q}_i^{1/2}\bs{Q}^{-1/2}\right]_r\bs{Q}^{-1/2},
\end{equation}
where $[\cdot]_r$ denotes the matrix formed by selecting the $r$ dominant right singular vectors associated with the largest singular values.

We emphasize that the projection $\bf{G}_i$ in~\ref{eq:zproj} is constant across all subcarriers $n$, as it relies on long-term spatial statistics that are stable across the bandwidth. Consequently, while $\bf{G}_i$ performs dimensionality reduction and spatial interference suppression, it does not compensate for the frequency-selective small-scale fading. This is addressed by applying a per-subcarrier instantaneous \ac{MMSE} equalizer $\bf{W}_i[n,k]$ after obtaining the projected signal $\bf{z}_i[n,k]$:
\begin{equation}
\wh{\bs{x}}_i[n,k] = \bs{W}_i[n,k] \bs{z}_i[n,k],
\end{equation}
where $\bs{W}_i[n,k]$ is typically a standard \ac{MMSE} or \ac{ZF} equalizer computed based on $\tilde{\bs{H}}_i[n,k] = \bs{G}_i \bs{H}_i[n,k]$. Because the dimension of $\bs{z}_i$ ($r$) is much smaller than the antenna dimension ($N_{rx}$), calculating $\bs{W}_i[n,k]$ is computationally inexpensive despite being performed for every subcarrier.

\textbf{Matrix Inverse Approximation for LTBF:} 
To reduce computational complexity and facilitate efficient hardware implementation, we adopt hardware-friendly matrix inversion approximation techniques to compute the inverse in \eqref{eq:Giopt}. It is important to note that in the literature on instantaneous \ac{MMSE} detection, the matrix $\bf{H}^H \bf{H} + \bf{I}$ (i.e., the Gram matrix) is often diagonally dominant and sparse, which enables the efficient use of various iterative and preconditioning methods~\cite{xue2016fast, albreem2021low, fang2025finite}. However, these properties do not generally hold for $\bf{H}\bf{H}^H + \bf{I}$, or matrix $\bf{Q}$. Nevertheless, $\bf{H}\bf{H}^H$ is a low-rank matrix, and its non-zero eigenvalues are identical to those of $\bf{H}^H \bf{H}$. As a consequence, $N_{\mathrm{rx}} - \sum_i N_s$ eigenvalues of $\bf{Q}$ are clustered around 1, while the remaining eigenvalues are relatively larger.

 \section{Interference Mitigation in Long-term Beamforming}

We study the impact of strong interference on \ac{LTBF} when low-complexity matrix inversion approximation methods are applied. We then introduce a subspace nulling approach based on long-term channel statistics to mitigate its effects.

\subsection{Impact of Strong Interference on Inverse Approximations}

As the relative interference power increases, the dynamic range of the received signal also increases, leading to a larger eigenvalue spread of the covariance matrix $\bs{Q}$. This, in turn, degrades the accuracy of matrix inverse approximations in \eqref{eq:Giopt}. To illustrate this effect, we consider two representative approaches: a non-iterative method based on polynomial approximation and an iterative method based on the \ac{CG} algorithm.

\textbf{Conjugate Gradient (CG):} 
Among iterative methods, the \ac{CG} algorithm is particularly attractive due to its low computational complexity and hardware scalability~\cite{albreem2021low, fang2025finite}. Since the dominant operations in \ac{CG} are matrix-vector multiplications, it can be efficiently implemented in hardware using architectures such as systolic arrays~\cite{rasteh2025spatial}.

The convergence rate of the \ac{CG} method strongly depends on the eigenvalue distribution of $\mathbf{Q}$, and in particular on the clustering of its eigenvalues. As the dynamic range increases, the eigenvalue spread typically grows, which can be detrimental in finite-precision implementations. In such settings, rounding errors and numerical instability may arise, potentially destroying the orthogonality of the \ac{CG} search directions and leading to a significant residual gap~\cite{fang2025finite, greenbaum2021convergence}. 

\textbf{Polynomial approximation:} 
The \ac{CG} method depends on sequential vector operations that require specialized hardware like \acp{DSP} or vector units, limiting its efficiency on systolic arrays. In contrast, polynomial approximation enables fully centralized computation using dense matrix-matrix multiplications, making it better suited for systolic array architectures.

In polynomial approximation, an increase in the dynamic range leads to a larger eigenvalue spread of $\bs{Q}$, which, in turn, degrades the accuracy of polynomial-based matrix inverse approximations. Specifically, the approximation error grows with the condition number of $\bs{Q}$, requiring higher polynomial orders to achieve near-optimal performance~\cite{rasteh2025scalable}.

A detailed theoretical analysis of finite-precision effects is beyond the scope of this work.
We rely on simulation results to illustrate these phenomena.
A rigorous analysis, covering both floating-point and fixed-point implementations, is left for future work.




\subsection{Interference Nulling via Long-Term Statistics}
The key idea is to explicitly identify and suppress dominant interferers prior to constructing the \ac{LTBF} projection matrix using long-term channel statistics (nulling). Let $\bs{Q}_v$ be the estimated
spatial covariance matrix of a dominant interferer from Section~\ref{sec:sys-model}.  As discussed there,
this matrix can be either estimated from channel estimated (if the interferer sends a known reference signals) or non-coherently from null frames if there
is no reference.  

In general, the interferer will belong to a low-rank
subspace in any given long-term coherence period
since the long-term statistics of the channel do not significantly change.
We thus find a low-rank approximation:
\begin{equation} \label{eq:Qlowrank}
    \bs{Q}_v \approx \bs{H}_v\bs{H}_v\herm,
\end{equation}
where $\bs{H}_v$ is $N_{rx} \times q$ for some rank $q$.
We call $q$ the \emph{nulling rank}.




We then compute the orthogonal projector onto the null space of the interference subspace is given by
\begin{equation}\label{eq:Proj}
\bs{P}_v = \bs{I} - \bs{H}_v (\bs{H}_v\herm \bs{H}_v)^{-1} \bs{H}_v\herm.
\end{equation}
We then simply apply the projection to all incoming signals:
\begin{equation}
    \tilde{\bs{y}}[n,k] = \bs{P}_v \bs{y}[n,k],
\end{equation}
which attempts to null the interferer before further processing.  Standard LTBF can then be used on the projected signals.
Let
\begin{equation}
    \label{eq:RQ}
    \bs{R}_v = \bs{Q} - \alpha_v \bs{Q}_v,
\end{equation}
denote the interference-reduced covariance matrix. 
For example, applying the projection, the effective user covariance becomes
\begin{equation}\label{eq:bigQ}
    \hat{\bs{Q}}_i = \bs{P}_v \bs{Q}_i \bs{P}_v^H.
\end{equation}
The resulting \ac{LTBF} projection matrix is then given by
\begin{equation} \label{eq:Giopt_nulled}
    \bs{\hat G}_i = \left[\hat{\bs{Q}}_i^{1/2}\bs{R}_v^{-1/2}\right]_r \bs{R}_v^{-1/2},
\end{equation}
which replaces $\bs{G}_i$ in \eqref{eq:zproj}. The receiver thus performs long-term interference suppression followed by low-rank beamforming in the interference-reduced subspace, while maintaining low computational complexity and relying only on long-term statistics.


\subsection{Computational Complexity}


A direct computation of \eqref{eq:Giopt_nulled} requires $\mathcal{O}(N_{\mathrm{rx}}^3)$ operations. To reduce complexity, we exploit the structure of the projection matrix in \eqref{eq:Proj}. Let
\begin{equation}
\bs{M} = (\bs{H}_v^{H} \bs{H}_v)^{-1},
\end{equation}
whose computation has complexity $\mathcal{O}(q^3)$ and is negligible for $q \ll N_{\mathrm{rx}}$. Substituting \eqref{eq:Proj} into $\hat{\bs{Q}}_i$ yields
\begin{align}
\hat{\bs{Q}}_i 
&= (\bs{I} - \bs{H}_v \bs{M} \bs{H}_v^{H}) \bs{Q}_i (\bs{I} - \bs{H}_v \bs{M} \bs{H}_v^{H})^{H} \\
&= \bs{Q}_i 
- \bs{H}_v \bs{M} \bs{H}_v^{H} \bs{Q}_i 
- \bs{Q}_i \bs{H}_v \bs{M} \bs{H}_v^{H} \nonumber \\
&\quad + \bs{H}_v \bs{M} \bs{H}_v^{H} \bs{Q}_i \bs{H}_v \bs{M} \bs{H}_v^{H}.
\label{eq:extend}
\end{align}

The dominant computational steps in \eqref{eq:extend} are the products
\begin{equation}
\bs{Q}_i \bs{H}_v \in \mathbb{C}^{N_{\mathrm{rx}} \times q}, \quad
\bs{H}_v^{H} \bs{Q}_i \in \mathbb{C}^{q \times N_{\mathrm{rx}}},
\end{equation}
each requiring $\mathcal{O}(N_{\mathrm{rx}}^2 q)$ operations. All remaining terms involve matrices of size at most $q \times q$ or $N_{\mathrm{rx}} \times q$ and therefore contribute lower-order complexity.

\begin{table}[!t]
\centering
\scriptsize
\setlength{\tabcolsep}{3pt}
\renewcommand{\arraystretch}{0.95}
\caption{Computational complexity (in complex operations) of LTBF with and without interference nulling ($N=N_{\mathrm{rx}}$).}
\label{tab:complexity}
\resizebox{\columnwidth}{!}{%
\begin{tabular}{@{}lcc@{}}
\toprule
\multirow{2}{*}{\textbf{Operation}} & \multicolumn{2}{c}{\textbf{Complexity}} \\
\cmidrule(l){2-3}
& \textbf{With nulling} & \textbf{Without} \\
\midrule
$\bs{Q}_j$ \eqref{eq:srs}
& $\mathcal{O}(N^2 N_{\mathrm{srs}} N_{\mathrm{UE}})$
& $\mathcal{O}(N^2 N_{\mathrm{srs}} N_{\mathrm{UE}})$ \\

$\hat{\bs{Q}}_j$ \eqref{eq:extend}
& $\mathcal{O}(q N^2 N_{\mathrm{UE}})$
& -- \\

$\bs{R}_v$ \eqref{eq:RQ} / $\bs{Q}$ \eqref{eq:Qdef}
& $\mathcal{O}(N^2 N_{\mathrm{UE}})$
& $\mathcal{O}(N^2 N_{\mathrm{UE}})$ \\

CG inverse
& $\mathcal{O}(k'N^3)$
& $\mathcal{O}(kN^3)$ \\

Poly. inverse
& $\mathcal{O}((d'-1)N^3)$
& $\mathcal{O}((d-1)N^3)$ \\

$\hat{\bs{G}}_i$ \eqref{eq:Giopt_nulled} / $\bs{G}_i$ \eqref{eq:Giopt}
& $\mathcal{O}(N^{2} (N_{\mathrm{srs}}+r) N_{\subsf \mathrm{UE}})$
& $\mathcal{O}(N^{2} (N_{\mathrm{srs}}+r) N_{\subsf \mathrm{UE}})$ \\

Projection \eqref{eq:zproj}
& $\mathcal{O}(r N N_{\subsf \mathrm{UE}})$
& $\mathcal{O}(r N N_{\subsf \mathrm{UE}})$ \\
\bottomrule
\end{tabular}%
}
\end{table}

Table~\ref{tab:complexity} summarizes the overall computational cost of inverse approximation methods, with and without interference suppression. Interference nulling introduces a one-time computational cost of $\mathcal{O}(q\,N_{rx}^2)$ per UE, corresponding to $\mathcal{O}(q\,N_{rx}^2\,N_{\mathrm{UE}})$ overall. However, as $N_{rx}$ becomes large, this overhead is negligible compared to the computational savings enabled by faster convergence.

In particular, interference nulling improves the conditioning and eigenvalue distribution of the covariance matrix. This leads to accelerated convergence of iterative or polynomial matrix inversion methods, which are performed once for all UEs. Consequently, the required number of CG iterations is reduced from $k$ to $k' \ll k$, or equivalently, the polynomial order decreases from $d$ to $d' \ll d$, as confirmed by the simulations.

 \section{Ray-Tracing Simulation Results} \label{results}

We evaluate the impact of a dominant interferer and the effectiveness of the proposed interference suppression on matrix inversion approximation methods. Results are presented under four numerical precision regimes:
\begin{itemize}
    \item \textbf{FP64} (double precision floating point),
    \item \textbf{FP32} (single precision floating point),
    \item \textbf{Q15.16} (fixed-point, 32-bit word, 16 fractional bits), 
    \item \textbf{Q7.16} (fixed-point, 24-bit word, 16 fractional bits).
\end{itemize}

In all scenarios, a \SI{30}{dB} interferer is introduced. The performance with the proposed nulling method is compared against the case without nulling prior to \ac{LTBF}. In the figures, the optimal \ac{MMSE} beamformer \cite{rasteh2025scalable} and exact \ac{LTBF} (\ac{LTBF} in which matrix inversion is computed without any approximation) are included as baselines, both computed in FP64.

\subsection{Simulation Setup}

We evaluate the proposed method using ray-tracing simulations based on the NVIDIA Sionna ~\cite{hoydis2023sionna} ray tracer. The scenario consists of a single \ac{BS} with three sectors, each equipped with a $16 \times 16$ antenna array and operating at a carrier frequency of \SI{3.5}{GHz}. The transmit power of each user is adjusted such that the post-beamforming \ac{SNR} lies within the range of $[-6, 14]$~dB, while the interference power is fixed at \SI{30}{dB}. The remaining parameters follow the 5G \ac{NR} specifications and are consistent with~\cite[Table~II]{rasteh2025scalable}.

Channel estimation is performed using \ac{SRS} over a long-term estimation window of $T_{\mathrm{LT}} = 10$~ms. The spatial covariance matrices and \ac{LTBF} projections are computed at the beginning of this interval, while performance (SINR) is evaluated at the end of the interval to capture the effect of channel evolution due to mobility. This setup provides a fair comparison with instantaneous \ac{MMSE} beamforming.

\subsection{Numerical Results and Discussion}
\begin{figure}[!t]
\centering

\begin{subfigure}{0.35\textwidth}
    \centering
    \includegraphics[width=\linewidth]{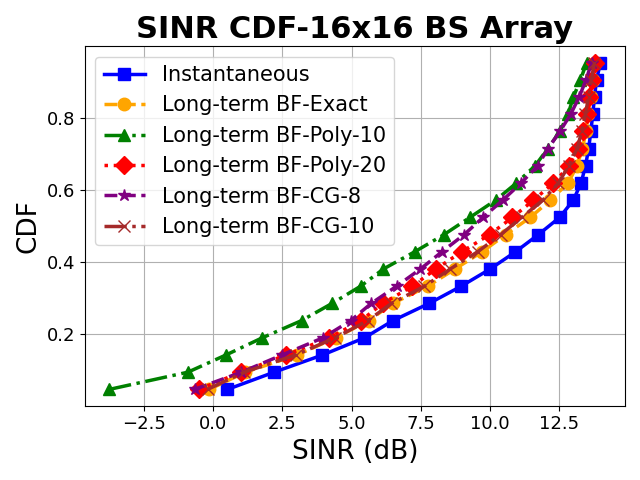}
    \caption{With Nulling}
    \label{fig:30db_nulling}
\end{subfigure}\hfill
\begin{subfigure}{0.35\textwidth}
    \centering
    \includegraphics[width=\linewidth]{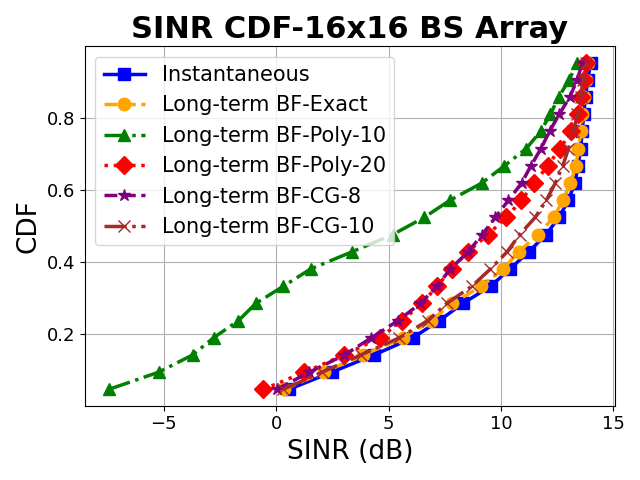}
    \caption{Without Nulling}
    \label{fig:30db_withoutnulling}
\end{subfigure}

\caption{Cumulative distribution function (CDF) of post-beamforming SINR under a 30 dB interferer in \textbf{FP64} precision, (a) with and (b) without nulling.}
\label{fig:sinr_cdf_-6-14}

\end{figure}

Fig.~\ref{fig:sinr_cdf_-6-14} shows that, under FP64 precision, increasing interference power necessitates higher polynomial orders to maintain accurate matrix inversion and near-optimal performance. The proposed nulling approach significantly reduces the performance gap between exact \ac{LTBF} and its approximations, indicating improved numerical accuracy and a lower required polynomial order. In contrast, the \ac{CG} method demonstrates greater robustness to interference, achieving performance close to exact \ac{LTBF} with a limited number of iterations. Additionally, nulling introduces a small performance gap between \ac{LTBF} and the instantaneous \ac{MMSE} benchmark. This is inherent to the approach, as the projection relies on long-term statistics that cannot fully capture instantaneous channel variations, leading to a slight loss of optimality.

\begin{figure}
    \centering
    \includegraphics[width=0.86\linewidth]{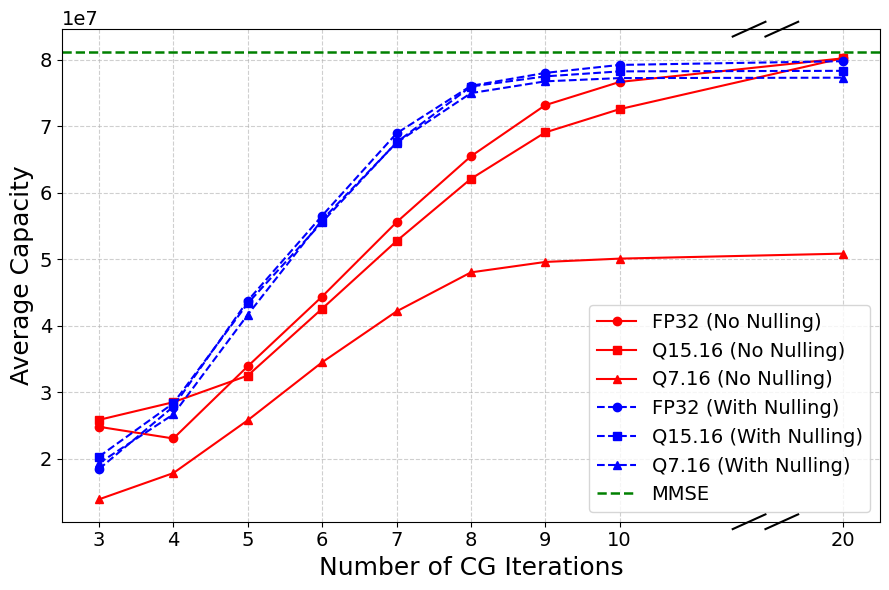}
    \caption{Average capacity versus number of conjugate gradient (CG) iterations, with and without interference nulling.
}
    \label{fig:cap}
\end{figure}

Focusing on the \ac{CG} method, Fig.~\ref{fig:cap} illustrates the impact of nulling across lower numerical precision regimes as a function of the number of iterations. Nulling significantly reduces sensitivity to numerical precision, enabling all implementations to converge close to the \ac{MMSE} benchmark. In contrast, without nulling, finite-precision effects become pronounced and limit achievable performance.

For FP32 and Q15.16, nulling substantially reduces the number of iterations required to reach near-optimal performance, requiring approximately 2 and 3 iterations, respectively. The most critical behavior is observed for the Q7.16 format, which exhibits slow convergence and a clear error floor due to numerical errors. In this case, increasing the number of iterations does not lead to near-optimal performance without nulling, highlighting the severe limitations imposed by low numerical precision in ill-conditioned scenarios.

\begin{figure}
    \centering
    \includegraphics[width=0.91\linewidth]{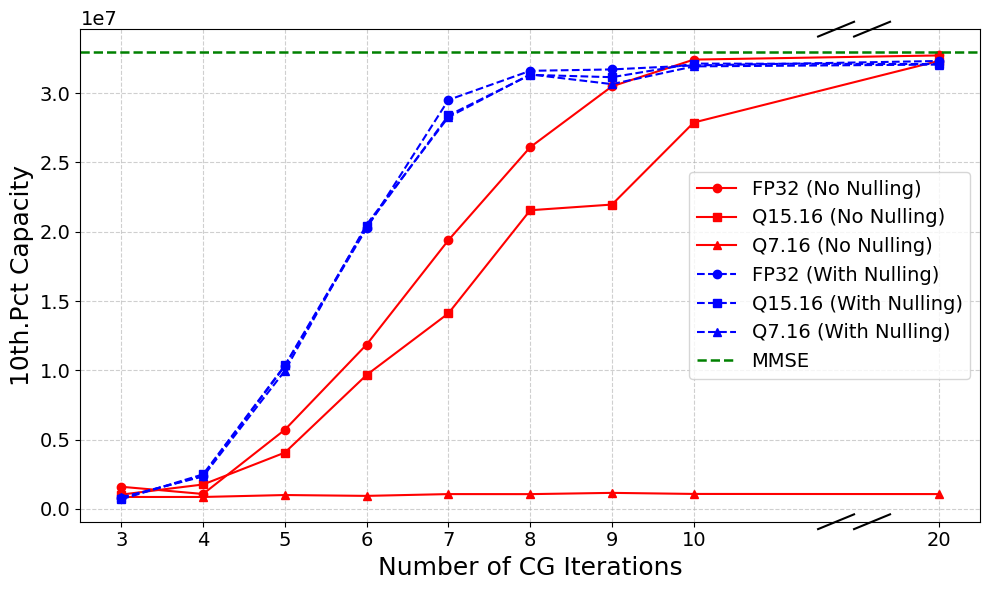}
    \caption{10th percentile capacity versus number of CG iterations, with and without interference nulling.}
    \label{fig:cap10th}
\end{figure}

Fig.~\ref{fig:cap10th} illustrates the impact of nulling on the 10th percentile capacity, a key metric characterizing cell-edge user performance, across different numerical precision regimes. Compared to the average capacity, the benefits of nulling are even more pronounced in this performance metric. Nulling significantly improves robustness, enabling all precision formats to approach the \ac{MMSE} benchmark with substantially fewer iterations. Furthermore, the most pronounced impact can be observed for the Q7.16 format. In this case, the 10th percentile capacity is significantly degraded, and almost no improvement is observed as the number of iterations increases. With nulling, however, Q7.16 exhibits a substantial performance recovery, closely tracking higher-precision formats. {This highlights that nulling is particularly critical for ensuring reliable performance in the worst-case user conditions and low precision implementations, where numerical errors and interference effects are most detrimental.} Overall, these results confirm that interference nulling improves robustness to finite-precision effects by mitigating ill-conditioning, thereby enabling more reliable convergence of the \ac{CG} algorithm. 

 \section{Conclusion}
In this work, we investigated the impact of strong interference on long-term beamforming (LTBF) under hardware-friendly matrix inversion approximations. We showed that dominant interferers significantly degrade the numerical conditioning of the covariance matrix, leading to severe performance loss, particularly in low-precision implementations. To address this issue, we proposed a subspace nulling approach based on long-term channel statistics. The method effectively reduces the eigenvalue spread of the covariance matrix, acting as an implicit preconditioning step that improves numerical stability. Ray-tracing simulation results demonstrated that the proposed approach substantially reduces the number of iterations required by conjugate gradient and polynomial methods to achieve near-optimal performance across floating-point and fixed-point implementations. Overall, interference-aware nulling enables robust and efficient LTBF operation in practical systems, with minimal additional complexity. Future work will focus on the analytical characterization of finite-precision effects and extensions to more dynamic interference scenarios.


\bibliographystyle{refs/IEEEtran}
\bibliography{refs/IEEEfull,refs/bibliography}{}


\end{document}